# An In-depth Summary of Recent Artificial Intelligence Applications in Drug Design

Yi Zhang[1]

## ABSTRACT

As a promising tool to navigate in the vast chemical space, artificial intelligence (AI) is leveraged for drug design. From the year 2017 to 2021, the number of applications of several recent AI models (i.e. graph neural network (GNN), recurrent neural network (RNN), variation autoencoder (VAE), generative adversarial network (GAN), flow and reinforcement learning (RL)) in drug design increases significantly. Many relevant literature reviews exist. However, none of them provides an **in-depth** summary of **many** applications of the recent AI models in drug design. To complement the existing literature, this survey includes the theoretical development of the previously mentioned AI models and detailed summaries of 42 recent applications of AI in drug design. Concretely, 13 of them leverage GNN for molecular property prediction and 29 of them use RL and/or deep generative models for molecule generation and optimization. In most cases, the focus of the summary is the models, their variants, and modifications for specific tasks in drug design. Moreover, 60 additional applications of AI in molecule generation and optimization are briefly summarized in a table. Finally, this survey provides a holistic discussion of the abundant applications so that the tasks, potential solutions, and challenges in AI-based drug design become evident.

[1]University of Minnesota – Twin Cities, Department of Computer Science and Engineering. Correspondence to: Yi Zhang <zhan2854@umn.edu>

## 1. INTRODUCTION

**Background.** The research field of molecular design by artificially intelligence (AI) has received much attention primarily because AI is perhaps by far the most promising approach to effectively explore the molecular space that is way too vast for human intelligence to handle. The field of AI-based molecular design can be divided to subfields based on molecule categories such as small molecules, biomolecules (e.g. protein, RNA and DNA), and inorganic molecules (Chen et al., 2020a; Eismann et al., 2021; Ingraham et al., 2019; Jing et al., 2021; Xiong et al., 2021; Yao et al., 2021). Among these subfields, AI-based design of small molecule is the subject of this survey, although the approaches from different subfields can often be used interchangeably with minor modifications such as changing the molecular representation (Jiang et al., 2020). One common purpose of small molecules is to act as small molecule drugs for treating diseases. To be succinct, the terms "drug" and "molecule" mean "small molecule drug" and "small molecule" in this survey unless otherwise specified.

Drug design is a part of the lengthy and costly drug discovery and development process (Blass, 2015; Wong et al., 2018). The entire process generally consists with four stages: 1. disease target identification, 2. molecule screening and lead discovery, 3. preclinical development involving animal testing, and 4. clinical development involving human testing (Tonkens, 2005). AI-based drug design has the potential to greatly simplify the stage two. Concretely, it could identify or generate molecular structures that are effective toward the disease targets and possess other drug properties. As a result, AI could expedite the drug discovery and development process and increase its return of investment. In addition, AI-based drug design methods can be facilely incorporated into laboratory automation, which is another field receiving much research interest (Coley et al., 2020a; Coley et al., 2020b; Dimitrov et al., 2019).

The applications of AI in drug design are centered around deep learning (DL) recently. The increased popularity of DL in drug design can be attributed primarily to three reasons. The obvious one is the theoretical advancement in deep learning. A large part of the theories related to graph neural network (GNN) and deep generative models were established over the past 10 years (Bronstein et al., 2017; Guo & Zhao, 2020; Liu & Zhou, 2020). New models and their variants are being rapidly leveraged or developed to solve challenging tasks in drug design by the efforts from both the computer science community and the chemistry community. Another reason for the increased use of DL in drug design is the increased molecular data availability. Much data is required to train a deep neural network with many parameters. Currently, there exist several large public chemical datasets that contain information about general molecular attributes (Gaulton et al., 2012; Kim et al., 2016). In addition, the recently improved high-throughput screening technologies can experimentally acquire molecular properties of interest in an unprecedentedly fast pace (Baudis et al., 2014; David et al., 2019). However, the author needs to caution the readers that in many cases, low data availability and quality are still the culprit for the subpar performance of DL models used in drug design (Walters & Barzilay, 2021). The third reason for the popularity of DL in drug design is the advancement of hardware. Compared to the conventional CPU, GPU and TPU train deep neural networks much faster (Jouppi et al., 2017). In addition, cloud computing renders GPU and TPU accessible to the users who don't own the necessary hardware (Armbrust et al., 2010).

AI-based drug design has three common purposes: molecular property prediction, molecule generation and molecule optimization. Molecular property prediction is mostly to determine the quantitative structure-activity relationship (QSAR). Prior to the emergence of DL, other machine learning (ML) techniques and computation models were used for predicting QSAR (Chen et al., 2018; Zhang et al., 2017). When using general ML including DL, predicting molecular property is essentially regression if continuous values of molecular properties are predicted, or, it is classification if discrete class labels are predicted. In DL, depending on the choices of molecular representation and other considerations, convolutional neural network (CNN), recurrent neural network (RNN), graph neural network (GNN) and/or multi-layer perceptron (MLP) can be used for molecular property prediction. DL-based molecular property prediction has achieved at least two real-world milestones. Stokes et al. (2020) identified halicin, an antibiotic effective in treating infections in animal model, from a compound library via GNN. Sakai et al. (2021) identified a highly active serotonin transporter inhibitor with antidepressant effects in behavior studies by GNN. Besides achieving great results by itself, DL-based molecular property prediction is often a part of molecule generation, which is discussed next.

Here, molecule generation is also called inverse QSAR or de-novo (i.e. from the beginning) molecule design. Molecular property prediction and molecule generation share the similar purpose to output molecules with properties of interest. However, molecules with the desired properties may not be in the existing molecular database and the unexplored chemical space is extremely vast. The motivation of designing techniques for molecule generation is mostly to navigate in the unexplored chemical space. Computational molecule generation is not new, and methods that are not based on DL were frequently used previously (Schneider & Fechner, 2005). Deep generative models, originally developed for other fields such as natural language processing (NLP) and computer vision, have been dominant recently in the field of molecule generation (Oussidi & Elhassouny, 2018). Deep generative models are often based on RNN, variational autoencoder (VAE), generative adversarial network (GAN) and flow.

The third subfield of drug design is molecule optimization, which aims to improve the properties of the molecules. Molecule optimization is often performed with molecule generation concurrently via reinforcement learning (RL), Bayesian optimization (BO) and other approaches (Frazier, 2018; Gómez-Bombarelli et al., 2018; Popova et al., 2019). It is considered separately here because there are AI-based models designed to specifically modify molecules for property improvement, rather than generating molecules from scratch (Zhou et al., 2019). Such models could be particularly useful in the lead optimization step of the drug discovery process.

The background part of this survey ends with a brief discussion of the molecular representations for AI-based drug design. The most frequently mentioned representations in this survey are Simplified Molecular Input Line Entry System (SMILES) and molecular graph. SMILES molecular representation, developed in 1988, is based on text strings (Weininger, 1988). Text string representation renders many NLP techniques readily appliable to drug design. However, SMILES is not a natural representation of molecules. This motivates the use of molecular graph, a more interpretable molecular representation (David et al., 2020; Gaudelet et al., 2020). In most cases, molecular graphs are constructed by using atoms or substructures as nodes and chemical bonds as edges.

**Table 1. Literature reviews with topics pertinent to this survey.**

| Citations | Topics and/or highlights |
|---|---|
| Bian et al., 2021<br>Deng et al., 2021<br>Kell et al., 2020 | Overviews of theory, implementation and examples of DL and generative models in molecular design. |
| Walters & Barzilay, 2021 | An overview of applications of AI in areas pertinent to drug discovery (e.g. molecular design, organic synthesis, and image analysis). |
| Schneider et al., 2020 | A discussion of challenges and possible solutions in AI-based drug discovery. |
| Walters & Barzilay, 2020 | A discussion of challenges and practical considerations in molecular property prediction and molecule generation aided by DL. |
| Xue et al., 2019 | A discussion of advances and challenges of molecule generation via DL. |
| Chen et al., 2018<br>Elton et al., 2019 | Early reviews of molecular design aided by DL. |
| Yang et al., 2019b | A broad overview of concepts and applications of AI in drug discovery before 2019. |
| Sun et al., 2020<br>Xiong et al., 2021 | Discussions of GNN applications in molecular design and synthesis. |
| Wieder et al., 2020 | A comprehensive enumeration of GNN applications in molecular property prediction by 2020. |
| Brown et al., 2020 | A historic view and a practical consideration of the use of AI in molecular property prediction. |
| Alshehri et al., 2020 | An overview and a comparison of both knowledge-based and data-driven approaches for computational molecular design. |
| David et al., 2019 | Insights regarding the use of DL to process large-scale compound data in industrial pharmaceutical research. |
| Gantzer et al., 2020 | An overview of computational molecule generation models based on DL and many other techniques. |
| Jiménez-Luna et al., 2020 | A discussion of explainable AI in drug design. |
| Öztürk et al., 2020 | An overview of the literature that applies NLP techniques in drug discovery. |
| Vamathevan et al., 2019 | A broad overview of ML concepts and their applications in many pharmaceutical subfields. |
| Mater & Coote, 2019 | A broad overview of DL in chemistry including many aspects other than drug design. |
| Rifaioglu et al., 2019<br>Lavecchia, 2015 | Early overviews of general ML techniques, tools and data for molecular property prediction. |
| Zhang et al., 2017 | An overview and a historical perceptive of the transition from ML to DL in drug design. |
| Schneider & Fechner, 2005 | An overview of early computational molecule generation models before DL. |

The connectivity of a graph is often represented by the adjacency matrix, and the attributes of nodes and edges are often expressed by the feature tensors. Besides molecular graph, another graph type involved in this survey is the molecular network, which is constructed by using molecular entities as nodes and their interactions or similarities as edges. Compared to molecular graphs, molecular networks are typically much larger.

**Relevant literature reviews and my contributions.** The growing interest in AI-based drug design can be best manifested by numerous research papers published in this field within the last four years (i.e. 2017 – 2021). In addition, the author found 25 literature reviews with topics pertinent to this field. The author(s), year and topics/highlights of these reviews are summarized in Table 1. Although many of these reviews contribute significantly to the literature, none of them provides an in-depth summary of many applications of AI in drug design. The review by Xiong et al. (2021) provides detailed summaries of some, but not many AI applications in drug design. The review by Wieder et al. (2020) provides a comprehensive list of GNN applications in molecular property prediction; however, the description of each application contains few details. This survey is designed to complement the existing literature by providing in-depth summaries of a large number of applications of AI in drug design. Concretely, this survey summarizes 102 recent applications of AI in drug design. 42 of them (13 applications for molecular property prediction and 29 applications for molecule generation and optimization) are summarized with details in text. The remaining 60 applications (all about molecule generation and optimization) are summarized briefly in a table. More applications of AI in molecular property prediction can be found in the review by Wieder et al. (2020). In most cases, the focus of each summary in text is the theoretical model (e.g. VAE, GNN, GAN, etc.), their variants, and the modifications made to render these models/variants capable of handling specific tasks in drug design (e.g. molecule generation under the low-data scheme, 3-D molecule design, molecule generation given the disease context, molecule optimization with a fixed scaffold, generation of molecules with guaranteed synthesizability, etc.). This survey should be particularly helpful to the readers who have some background in DL and want to learn the details of AI-based drug design comprehensively and rapidly with the help of many examples.

**Survey organization.** Except Section 1 and 9, each section centers around one or more AI technique(s). In each of these sections, theoretical development of the AI technique(s) is first explained, and relevant applications follow the explanation of the theory one by one. Specifically, Section 2 covers the theory about perceptron, MLP, CNN, GNN, and their applications in molecular property prediction. Topics of Section 3 to 8 are RNN, RL, VAE, GAN, flow, and Mont Carlo tree search (MCTS) respectively. Applications for molecule generation and optimization (those that are summarized in the text) are included in these sections. Section 9 ends the survey with discussion, challenges, and conclusion about the AI-based drug design. Key abbreviations used in this survey are in Table 2.

**Table 2. Key abbreviations.**

| Full names | Abbreviations |
|---|---|
| Artificial intelligence | AI |
| Autoencoder | AE |
| Bayesian optimization | BO |
| Convolutional neural network | CNN |
| Deep learning | DL |
| Gated recurrent unit | GRU |
| Generative adversarial network | GAN |
| Graph neural network | GNN |
| Long short-term memory | LSTM |
| Machine learning | ML |
| Message passing neural network | MPNN |
| Mont Carlo tree search | MCTS |
| Multi-layer perceptron | MLP |
| Natural language processing | NLP |
| Quantitative structure-activity relationship | QSAR |
| Simplified molecular input line entry system | SMILES |
| Recurrent neural network | RNN |
| Reinforcement learning | RL |
| Variation autoencoder | VAE |

## 2. MOLECULAR PROPERTY PREDICTION VIA DEEP LEARNING

**Theories of perceptron, MLP, CNN, and GNN.** Perceptron is inspired by the biological neuron, and it is a one-layer feedforward neural network. Perceptron is considered as the elementary block of deep neural network. A perceptron can be used as a linear binary classifier outputting the classification predictions by multiplying and combining trainable weights and input features through the linear predictor function. Heaviside step function is used for activation. An adjustable bias term shifting the decision boundary is often used before activation to increase the classification accuracy. A perceptron can also be used for regression. During training, weights are initialized randomly and trained through gradient descent using each input sample separately. An epoch is finished when the model has been trained with all input samples for one time. Perceptron hyperparameters include the number of training epochs and the learning rate. Number of training epochs is typically associated with the stopping criteria. Too few and too many epochs lead to underfitting and overfitting respectively. The learning rate controls the learning speed of perceptron by affecting weight updates. These two hyperparameters are also important in general neural networks.

Multilayer perceptron is an extension of perceptron, and it is a feedforward neural network with at least one hidden layer in addition to the input and output layers (Pal & Mitra, 1992). Thus, MLP belongs to deep learning due to its multiple layers. Similarly, CNN, GNN, and RNN are also deep learning. Backpropagation is used to minimize the loss during MLP training. Note that the fully-connectivity of usual MLP could make it prone to overfitting.

Convolutional neural network is inspired by visual cortex, and it is highly suitable for grid-like inputs such as images (Albawi et al., 2017; Aloysius & Geetha, 2017). Thus, it is widely used in computer vision. Compared to MLP, CNN has fewer parameters, lower propensity to overfitting and higher efficiency; meanwhile, CNN can capture input spatial information. CNN has a complex architecture typically including an input layer, alternating convolution layers and pooling layers, fully connected layers, and an output layer. A convolution layer applies the dot product of the input matrix and the convolution filter (i.e. vector of weights and bias). ReLU is often used as the activation function. A feature map is completed after the path of the sliding filter covers the entire input matrix. For some task-related requirement and computational efficiency, pooling layers (e.g. max or average pooling) are used to reduce the intermediate data dimensions after convolution. Typically, fully connected layers similar to MLP are after the convolution and pooling layers. CNN architecture ends with the output layer. Compared to MLP, CNN has more hyperparameters such as filter size, filter stride, and pooling type.

**Theory of graph neural network.** MLP and CNN are designed specifically for the Euclidean data. Compared to Euclidean data, graphs are irregular due to varying numbers of nodes, and varying numbers of neighbor nodes of different nodes. Thus, conventional neural networks are not directly appliable to data in graph form. This motivates the development of GNN. Some pioneer works in GNN were done from 2005 to 2010 (Gallicchio & Micheli, 2010; Gori et al., 2005;

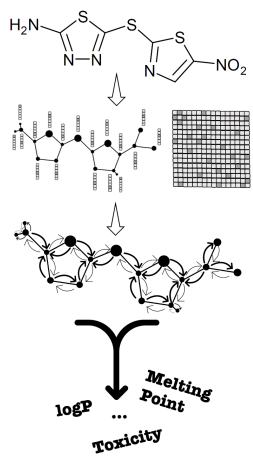

**Figure 1. A simplified example of using ConvGNN for graph classification/regression in molecular property prediction.** Each molecule is represented by a graph with feature vectors and an adjacency matrix. Message exchange updates the node information. The graph embedding is then obtained and used for property prediction.

Scarselli et al., 2008). Numerous GNN variants are proposed afterwards, and they can be classified into three categories: recurrent GNN (RecGNN), convolutional GNN (ConvGNN), and spatial-temporal GNN (STGNN). I mainly discuss the applications of ConvGNN in this survey. Applications of RecGNN in molecular property prediction can be found in papers by Lusci et al. (2013) and Wieder et al. (2020). STGNN applies to graphs of which node attributes vary over time, thus, it is irrelevant to this survey.

ConvGNN can be further divided into two categories: spectral ConvGNN and spatial ConvGNN. Spectral ConvGNN uses a spectral filter on spectral parts after decomposing graph signal on the spectral domain. ChebNet is proposed to reduce the computation cost associated with spectral ConvGNN by circumventing the computation of the Laplacian eigenvectors (Defferrard et al., 2016). In ChebNet, the spectral filter is approximated by Chebyshev polynomial. Graph convolutional network (GCN) is proposed to simplify ChebNet by using more approximations (Kipf & Welling, 2016). GCN bridges the spectral and spatial ConvGNNs. Spatial ConvGNN updates the feature of each node by convolving its current feature with the features from its neighboring nodes. Spatial ConvGNN is the focus of this review, its details will be illustrated later through a series of examples.

ConvGNN is mainly used for three tasks: node classification, link prediction, and graph classification/regression. Node classification typically belongs to the semi-supervised learning category, and it is for predicting the labels of unlabeled nodes in a graph given the labels of other nodes. The link prediction task is for predicting the presence and identity of the edge when the information of two nodes is given. Link prediction has been applied to predict drug-target and drug-drug interactions when the graph is a molecular network (Huang et al., 2020). The central task of ConvGNN in drug design is graph classification/regression, which is used for molecular property prediction (Hwang et al., 2020). For graph classification/regression, a ConvGNN is often coupled with pooling and readout operations to obtain a compact graph representation, followed by using a MLP to complete the end-to end framework. A simplified example of using ConvGNN for graph classification/regression is in Figure 1. More details for GNN can be found elsewhere (Wu et al., 2020; Zhang et al., 2019b; Zhou et al., 2020). The remaining text of Section 2 summarizes 13 applications/developments of GNN for molecular property prediction. Note that the main subject of the last three examples is link prediction of molecular network.

**Work by Duvenaud et al. (2015).** Duvenaud et al. develop a pioneer spatial ConvGNN model for molecular property prediction by modifying the circular fingerprint model, which is one of the state-of-art models at that time (Glen et al., 2006). In the circular fingerprint model, feature vectors associated with the molecules are computed by the circular fingerprint algorithm. The feature vectors are subsequently fed to a fully connected neural network for chemical property prediction. The circular fingerprint algorithm is a good starting point for developing neural network suitable for graphs partly because circular fingerprint is similar to CNN. Concretely, circular fingerprint method uses the same local operation everywhere, followed by a pooling step. However, circular fingerprint is not differentiable due to the hash and indexing operations. Consequently, circular fingerprint model cannot be trained end-to-end. To overcome this issue, the hash operation is replaced by a one-layer neural network and the indexing operation is replaced by SoftMax. After modifications, an early spatial ConvGNN model allowing end-to-end training is developed for molecular property prediction and it can operate directly on molecular graphs with arbitrary sizes and shapes.

**Work by Coley et al. (2017).** Most chemical properties (e.g. logP and melting point) are heavily impacted by the molecule global information beyond connectivity. To improve the prediction accuracy of such chemical properties, Coley et al. proposed another spatial ConvGNN model related to the model by Duvenaud et al. Their strategy to help the model capture molecule global information is to incorporate the "global property contribution" of an atom to the associated atom features. For example, the total polar surface area of a molecule is collectively determined by all the atoms. Thus, the polar surface area contribution from each atom can be used as part of the associated atom feature. At each layer/depth of their GNN model, the features of each atom are updated based on the feature vectors of the atom itself, its neighboring nodes, and the edges. The edge features remain unchanged. The updated atom feature vectors of all atoms are converted to atom fingerprints, which are summed to form the molecular fingerprint. The process is repeated for each layer of GNN. Finally, the molecular fingerprints obtained from all layers are summed and then used to predict the molecule property by the downstream neural network.

In addition, the authors point out the importance of molecular embedding strategy in addition to the downstream regression. Conventional descriptor-based molecular embedding could have subpar performance if the descriptors omit key molecular information. By contrast, the GNN embedding approach renders the complicate QSAR learnable directly from the molecular graph and the features.

**D-MPNN (directed message passing neural network) (Yang et al., 2019a).** D-MPNN is based on the previous model by Dai et al. (2016). Part of the model's name, MPNN, is from the general MPNN framework proposed by Gilmer et al. (2017). The graph convolution of many GNN models is typically node-centered. By contrast, the convolution of D-MPNN is centered on the directed edges. The authors point out that such modification can prevent totters associated with node-centered message passing and thus reduce noise associated with the graph representation. After the edge-based message passing, the node representations are updated based on the incoming edge features. The readout phase uses the node representations to determine the molecule embedding, which is subsequently used for property prediction. The authors also propose a strategy to help the molecule embedding capture the global molecular information. The global molecular features are first computed via RDKit, a common cheminformatics software (Landrum, 2013). During the readout phase of D-MPNN, the global molecular features are concatenated with the learned molecule embedding by D-MPNN for property prediction.

**Identification of halicin (Stokes et al., 2020).** D-MPNN is the model used to identify halicin as an antibacterial candidate that shows effectiveness in treating infections in murine model. In this groundbreaking work, D-MPNN is trained with 2,335 molecules to predict the antibacterial property of given molecules. The trained model is subsequently used to predict the antibacterial properties of >107 million molecules from multiple libraries. Compounds with promising antibacterial properties, divergent structures, and high availabilities are selected in priority. This procedure ultimately leads to the identification of halicin from the Drug Repurposing Hub. In addition, halicin shows divergent structure when compared to the conventional antibiotics. This indicates the generalizability of the model to predict properties of new chemicals.

**MPNN with an extra global node (Li et al., 2017).** Li et al. adopt another approach to enable GNN to capture global molecular information. Before introducing their model, the authors explain why vanilla MPNN fails to capture molecule global information. Vanilla MPNN typically contains few layers, thus the resulted receptive field is small when compared with the molecule size. The receptive field can't be simply enlarged by adding more layers to capture global molecular information because more layers can cause overfitting given the limited amount of training data.

In their model, Li et al. add an extra global node to the molecular graph. The global node is connected to all local nodes in the graph via directed edges pointing from the local nodes to the global node. During message passing, the information of local nodes is updated. Meanwhile, due to the directed edges toward the global node, the feature of the global node is also updated using the features of all local nodes. As a result, the global node can capture the information of the entire graph through as few as one layer of message passing. The feature of the global node is used as the molecular feature for the downstream classification or regression.

**Hierarchical inter-message passing (Fey et al., 2020).** Fey et al. propose a hierarchical inter-message passing model to capture molecular hierarchical information such as ring structure. Their model operates on two graphs: the raw molecular graph and its associated junction tree (JT). The nodes of JT are molecule substructures. The raw molecular graph and the JT capture fine details and coarse details of a molecule respectively. The concept of JT is first proposed in the JT-VAE model (Jin et al., 2018a), and it will be discussed in Section 5. Given coarse and fine graphs of a molecule, the molecule representation is learned by passing messages within each graph and between two graphs using coarse-to fine and fine-to-coarse information flows. As a result, molecular hierarchical information can be effectively represented.

**MPNN for tetrahedra chirality (Pattanaik et al., 2020).** An MPNN model is developed by Pattanaik et al. to handle molecules with tetrahedra chirality, one common type of stereochemistry. Molecules with different stereochemistry have the same graph connectivity; thus, stereochemistry can't be discerned by a conventional MPNN. However, stereochemistry impacts many molecule properties. The approach of including stereochemistry information into the feature vectors of the atoms or bonds is typically limited. The authors realize that the symmetric aggregation function (e.g. mean, sum, and max) is the part of conventional MPNN architecture that hampers the detection of stereochemistry. This motivates the authors to develop two aggregation functions for MPNN that can distinguish molecules with tetrahedra chirality in property prediction. They denote two aggregation functions as "Permutation" and "Permutation-concatenation". Both functions are incorporated into several GNN models (i.e. graph convolutional network, graph isomorphism network by Xu et al. (2018), and D-MPNN) to access performance. There are other stereochemistry types in addition to tetrahedral chirality. Developing GNN model for them will be an interesting direction.

**Pretraining GNN (Hu et al., 2019).** Pretraining is particularly useful when training and testing examples have different distributions and task-specific labels are insufficient. The state-of-the-art pre-training strategy such as graph-level multi-task supervised learning cannot significantly improve GNN predictive performance. Thus, the authors propose two self-supervised node-level pre-training techniques: context prediction and attribute masking. In context prediction, the substructure of a graph is used to predict the surrounding graph structure. In attribute masking, certain node/edge features are masked and GNN is used to predict these features according to the neighborhood information. This is an example of node classification/regression task. After node-level pretraining, the graph-level multi-task supervised learning is used for graph-level pretraining. Node-level pre-training is performed before graph-level pretraining to alleviate the negative impact of the irrelevant tasks in the graph-level multi-task supervised pretraining.

**MolGNN (predict COVID-19 drug candidates) (Liu et al., 2021b).** Given extremely few task-specific data, Liu et al. modify the model by Hu et al. to predict drug candidates for COVID-19 treatment. Specifically, Liu et al. modify the graph-level pre-training phase by using PubChem fingerprints to replace the commonly available molecule properties. They show that such modification improves the model performance. To obtain COVID-19-specific data for training the model after the pretraining phase, they use screening to find compounds that target enzymes associated with Covid-19 and select additional data from antiviral experiments.

**DGraphDTA (double graph drug target affinity prediction) (Jiang et al., 2020).** The aforementioned works belong to ligand-based scoring, which is to predict properties of drug based on the drug itself. A related but different category is drug-target affinity (DTA) prediction, which is an essential part of virtual screening. Examples of DTA metrics include binding affinity, dissociation constant, and IC50 value. In this example of DTA prediction enabled by GNN, the authors use two GNN models on the graph representations of both protein (target) and molecule (drug) to obtain two embeddings, which are concatenated for DTA prediction. The key contribution of this work is to address the challenges associated with protein graph construction. An unfavorably large graph is expected if a typical protein is treated as a small drug molecule during the graph construction. An alternative is to use amino acids as nodes and peptide bonds as edges to construct the graph. But this approach results in a long chain analogous to the protein primary structure; thus, the secondary and tertiary protein structures can't be effectively captured. To solve this issue, the authors use amino acids as nodes and a contact map to indicate the amino acid interactions. The contact map is built based on the protein sequence via the method Pconsc4 by Michel et al. (2019) and the resulting structure is equivalent to a graph.

**Decagon (Zitnik et al., 2018).** Starting from here, the examples are about using GNN for link prediction to predict general chemical interactions in molecule networks. As mentioned before, the network graphs are constructed by using chemical entities as nodes and interactions as edges. Decagon is proposed to predict the adverse side effects of drug combinations in human body due to drug-drug interaction. A graph is constructed using drugs and proteins as nodes. The nodes are linked based on three types of interaction: 1. protein-protein interaction, 2. drug-protein interaction, and 3. the side effect represented by drug-drug interaction. Each side effect is represented by an edge type. Thus, side effect prediction is transformed to edge identity prediction. The task in this work is characterized as multi-relational link prediction in multimodal networks. Decagon uses GNN as the encoder to embed drugs and proteins, and uses a tensor factorization model as the decoder to predict the side effects between drugs.

**DTI-GAT (drug target interaction-graph attention network) (Wang et al., 2021b).** DTI-GAT is used for drug-target interaction prediction. The graph representation in DTI-GAT resembles that of Decagon, where drug and protein are used as the nodes and there are multiple edge types. Specifically, the edge types of DTI-GAT are drug-drug similarity, protein-protein similarity, and drug-protein interaction. The authors point out that the popular bipartite graph representation using only drug-target interactions as edges limits the information flow during node embedding, and the use of heterogenous graph with more edges can alleviate this issue.

The authors adopt the graph attention network (Veličković et al., 2017), which uses self-attention mechanism that learns edge weight adaptively in a network. As a result, the prior knowledge about the relative importance of neighbor nodes for a given node is not required. After message passing, the embeddings of a pair of protein and drug are used for DTI prediction via a decoder. The authors also suggest that the learned attention weights can provide hints about the DTI topological structure and similarity.

**SkipGNN (Huang et al., 2020).** The authors of this work point out that many relevant works only capture the direct interactions (referred to as direct similarity) in a molecular network for molecular interaction prediction. The main contribution of SkipGNN is to capture the second-order interactions (referred to as skip similarity) in addition to the direct interactions. The molecular network is represented by two graphs: its raw graph for capturing direct similarity and a skip graph for capturing skip similarity. Two GNNs are applied to two network graphs during the node embedding phase and two GNNs are design to interact with each other. The obtained node embeddings are used for interaction prediction. The final model learns to balance the use of the direct similarity and skip similarity for interaction prediction.

# 3. MOLECULE GENERATION AND OPTIMIZATION BY RECURRENT NEURAL NETWORK

**Theory of RNN.** RNN, proposed in the 1980s (Hopfield, 1982; Jordan, 1986), is suitable to process data with sequence structure and data with different lengths. Sequences can be temporal or non-temporal, and this survey uses time step to denote step for sequences of both kinds. Unlike the standard feedforward neural network of which output only depends on the input, RNN determines the output at each time step by using the current input and the output from the previous time step. Generally, the final output is used to compute the loss, which is backpropagated through time steps in the reverse order (abbreviated as BPTT). Theoretically, long-range dependence in the input sequence can be tracked by a vanilla RNN. Yet, BPTT could cause the gradient to asymptotically approximate zero at very early time steps due to the involved computation of repeated matrix multiplication using finite precision numbers. Meanwhile, the gradient could also grow without bound. Two possibilities are termed as vanishing gradient and exploding gradient respectively (Pascanu et al., 2013).

Vanishing gradient problem prevents the gradient from reaching early time steps. As a result, vanilla RNN fails to track long-term dependence in practice. Long short-term memory (LSTM), proposed around 1997 (Hochreiter & Schmidhuber, 1997), is an effective architecture to alleviate the vanishing gradient issue and enable RNN to capture the long-range dependence. An example of LSTM unit has a cell and three gates (input gate, output gate, and forget gate). The cell remembers previous information, and gates regulate information flow. LSTM has complex architecture, and the involved computation is expensive. To address this issue, gated recurrent unit (GRU) was proposed in 2014 (Cho et al., 2014). Compared to LSTM, GRU requires fewer parameters and doesn't have the output gate; yet, it has similar performance. Another interesting RNN variant is bidirectional RNN, which also allows the information from future time steps, in addition to past time steps, to influence the output of the current time step (Schuster & Paliwal, 1997). RNN variants exist and more details about RNN are provided by Goldberg (2016), Lipton et al. (2015), and Tarwani & Edem (2017). RNN can be used for molecular property prediction by treating the molecules as sequences.

After being trained using the target sequences, RNN can also be used for generating new sequences according to the training set (Bacciu et al., 2020; Graves, 2013; Olivecrona et al., 2017). The generative process starts when the beginning of sequence is inputted. At each generative step, a probability distribution of the next sequence component is determined based on the intermediate sequence. One sequence component is sampled from the distribution and the intermediate sequence is updated. For example, an intermediate sequence could be a SMILES representation "C1CCC". During the generative process, different probabilities will be assigned to the next atom such as 'C', 'H', 'O', and "Cl". If 'C' is selected, the string becomes "C1CCCC" for the following generative step. The process repeats during the remaining generative process. The stochasticity of generative process leads to a variety of sequences. LSTM RNN can be used to generate complicate long-range sequence (Graves, 2013). RNN generative process often has convergence issue and instability. Teacher forcing is a common technique to address this issue (Kolen & Kremer, 2001). Instead of using the generated intermediate sequence to predict the next sequence component, teach forcing always use the ground truth sequence as the input for the next prediction. Here, only two applications of RNN in molecule generation and/or optimization are summarized. However, RNN is often used with other AI techniques, and more applications involving RNN will be provided in the future sections after these techniques are introduced.

**Work by Segler et al. (2018a).** Segler et al. develop an RNN-based generative model with three stacked LSTM layers and combine it with transfer learning to generate molecules active toward Staphylococcus aureus and Malaria under low data regime. Transfer learning is used to address the data insufficiency issue, which is common in AI-based drug design. In transfer learning, a large dataset is typically available to train the ML model for a relevant, but different task to learn the general features from the data, and then the model is retrained with a smaller data set for the task of interest. The process is called fine tuning. Transfer learning effectively prevents overfitting with a small data set (Zhuang et al., 2020). The authors first use a large dataset of general molecules represented by SMILES to train the RNN generative model, which can subsequently generate novel, but general molecules. Next, the model is fine-tuned with a small set of molecules active against Staphylococcus aureus and Malaria so that it can generate more molecules active to the same targets. In addition, the authors also simulate an example of automated drug discovery cycle with 6 steps: 1. molecule generation by RNN 2. virtual molecule synthesis 3. molecule property evaluation, 4. selection of molecules with desired properties, 5. retraining the RNN generative model using selected molecules and 6. Repetition.

**GraphRNN (You et al., 2018b).** GraphRNN is an RNN-based graph generation model that is trained on graphs. It is later modified and applied specifically for molecular graph generation process in MolecularRNN (Popova et al., 2019). During the inference stage of generation, GraphRNN decomposes the graph generation process into two coupled subprocesses: graph-level RNN and edge-level RNN. Graph-level RNN generates a sequence of nodes, and captures the graph state. For each newly added node in the sequence, a sequence of edges (represented by adjacency vector) is generated via the edge-level RNN which starts from the intermediate state of the graph-level RNN. Note that the GraphRNN generation process follows the breadth first order, which avoids training GraphRNN on all possible node permutations and reduces the number of edge predictions by the edge-level RNN. As a result, the scalability of GraphRNN is improved and larger graphs can be generated.

## 4. MOLECULE GENERATION AND OPTIMIZATION BY REINFORCEMENT LEARNING

**Theory of RL.** Dated back to 1950s, emergency of RL was inspired by behaviors of living entities. RL has a mix of deliberate planning and the "trial and error" approach. An agent (e.g. computer) has specific goals (e.g. designing drugs with high affinity and specificity toward certain disease target) and interacts with the environment (e.g. chemical environment such as electron density and conjugation) by evaluating the environment (i.e. obtain the state capturing the environment) and selecting actions from action space (e.g. changing chemical bond type and adding functional group) to impact the environment and lead to state transition. The action selection criterion based on the perceived state is policy, which ranges from simple lookup table to complicated and stochastic mappings. Each state transition results in a reward which is directly determined by state and the action. Agents aim to maximize the long-term cumulative reward. Reward indirectly depends on policy, which is subject to change to increase rewards. A long-term cumulative reward is collectively determined by the immediate reward of a state transition and the future reward of the subsequent state transitions. Clearly, exclusively maximizing immediate reward is short-sighted and could compromise the cumulative reward. For example, adding some functional groups that favor certain property to a molecule in early stages may ultimately lead to a subpar global structure compromising the property of interest. The concept related to far sight is value, which characterizes the expected reward that can be accumulated in the long run after a certain state. It is of particular interest to estimate value to maximize the cumulative reward. A concept closely related to value is Q-function, which mathematically expresses the expected accumulative reward given the current state and the action taken. Note that the tradeoff between exploiting experience and exploring new choices in the "trial and error" approach should be considered to maximize the total reward. Being tilted to either side compromises RL model and this concern motivates the adoption of strategies such as upper confidence bound algorithm, gradient bandit algorithms, and $\varepsilon$-greedy algorithms. A simplified example of RL in chemical environment is depicted in Figure 2. More RL details are provided by Arulkumaran et al. (2017), Kaelbling et al. (1996), and Moerland et al. (2020). Five applications centering around RL for molecule generation and optimization are summarized next. More applications involving RL in addition to other techniques such as VAE are summarized in future sections.

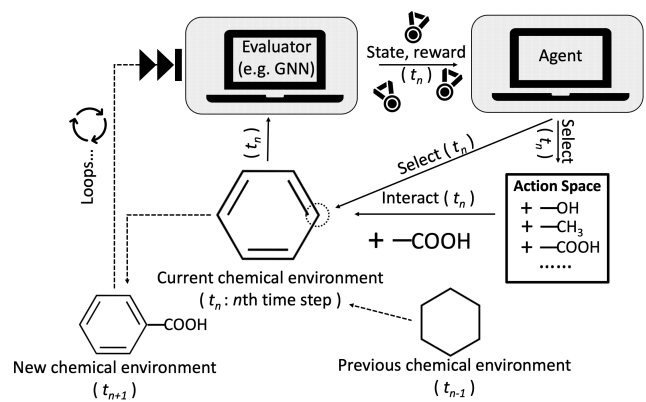

**Figure 2. A depiction of one possible setting of RL in chemical environment.**

**GCPN (graph convolutional policy network) (You et al., 2018a).** GCPN is an early RL-based molecule generative model trained by policy gradient. GCPN represents the molecules by graphs and uses GNN. The developers of GCPN prefer the graph representation to the SMILES representation for graph generation due to three reasons. First, a minor change in the text string representation can result in a significant change in its meaning. Second, text representation is incompatible with molecule valency check. Third, intermediate text string can be meaningless. Using graph representation circumvents three issues. GCPN includes a discriminator that is pre-trained with realistic drug molecules in a generative adversarial network framework, which is the topic of Section 6. After training, the domain knowledge from these molecules is implicitly captured by the discriminator all at once. The discriminator is included in the GCPN by using the adversarial loss as part of the reward of GCPN. Such design helps GCPN generate realistic molecules. The graph generation process is a sequential Markov decision process in which the next action/state depends only on the current state. The state in the graph generation process is an intermediate graph. The action in GCPN resembles link prediction task in GNN. Prior to an action, the intermediate graph is surrounded by nodes representing atoms. The action space includes connecting nodes within the intermediate graph, and connecting the intermediate graph with one surrounding node. After an action is taken, the disconnected nodes are removed. Higher intermediate reward is assigned if the modified graph is realistic and passes the valency check. When the molecule generation process terminates, a higher final reward is assigned if the generated molecule has desired properties and a stable structure.

**MolecularRNN (Popova et al., 2019).** MolecularRNN extends the aforementioned GraphRNN model to render it compatible with attributed molecular graphs. In addition, after pretraining, MolecularRNN is incorporated into an RL framework trained by policy gradient to generate molecules with desired properties. To generate attributed graphs, MolecularRNN predicts node and edge types. Specifically, the type of next node/atom is predicted after unrolling graph-level RNN across atoms. The edge/bond type is predicted by the edge-level RNN. Recall that GraphRNN uses adjacency vector with entries of 0/1 to indicate the presence or absence of bond at certain locations. In MolecularRNN, the 0/1 entries of the adjacency vector are replaced by numerical values to indicate bond types in addition to bond presence at different locations. The action space of MolecularRNN is to generate an atom of any type and connect it to any location of the intermediate graph via any edge type. Similar to GCPN, MolecularRNN performs valence check after each state transition to ensure validity. After termination, the final reward is assigned based on the desired property possessed by the generated molecule. The final reward is then distributed to all the intermediate steps.

**MolDQN (molecule deep Q-networks) (Zhou et al., 2019).** Compared to the aforementioned models, MolDQN has two major differences. First, MolDQN uses the value function learning method while GCPN and molecularRNN use the policy gradient method

 

(Mnih et al., 2015). Second, the task of MolDQN is molecule modification rather than molecule generation from the beginning. Thus, the starting state of MolDQN is a molecular graph. In their RL framework, the molecule modification action space includes both addition and deletion of bonds and atoms. The number of state transition can be used to partly control the extent of modification.

Molecule modification is often a crucial step in the drug discovery process after leads are identified. This process is part of lead optimization. Leads have desired central properties (e.g. effectiveness toward the disease targets), which are captured by their scaffolds. Yet, general drug properties such as solubility and logP of leads may be unsatisfactory. Thus, the central task in molecule modification is to improve the molecule general properties while maintaining the molecule central properties by keeping the scaffold. In this work, the authors adopt two methods, constrained optimization and multi-objective optimization, to accomplish this task. In their constrained optimization approach, the general properties of the molecule are optimized when the requirement of retaining the scaffold is used to construct the feasible region of the optimization problem. The second approach, the multi-objective optimization technique, which optimizes multiple objectives simultaneously to obtain the pareto-solution, is more flexible than constrained optimization in this application. To enable multi-objective optimization, the authors adopt the simple linear scalarization technique, which linearly combines multiple objective functions into a single one, and assigns weights to all objective components to regulate their relative importance. The objectives in their application are rewards associated with maintaining scaffolds (measured by Tanimoto similarity) and improving the general molecular properties. The weights associated with the rewards are varied, and molecules with a range of Tanimoto similarity and general properties are generated. No pre-training is used in this work as the authors argue that the dataset for pre-training may introduce bias in the modified molecules.

**3-D molecular design (Simm et al., 2020b).** Simm et al. developed an RL-based model for 3-D molecular design. Although graph representation is more suitable for molecular structure when compared to text string, graph representation still has several limitations. First, the graph representation is unsuitable for generating multi-molecule system. Second, using some physical laws for molecule property prediction requires molecular spatial information, which can't be conveniently captured by the 2-D graph representation. Third, geometric constraints are required during some molecule generation processes (e.g. generate molecules that fit a protein binding site), and graph representation is unsuitable for these tasks. To overcome these limitations of graph representation, the authors develop a model that generates molecules in a 3-D Cartesian coordinates. During each step of the generation process, an atom is selected from a list of options and placed to the 3-D canvas by an agent.

**REACTOR (reaction-driven objective reinforcement) (Horwood & Noutahi, 2020).** REACTOR is an RL model developed to address the poor synthesizability of molecules generated by many AI-based models. This issue is extensively discussed in the paper by Gao & Coley (2020). REACTOR uses the chemical reactions as the transitions during the Markov decision process for guided molecule optimization. As a result, each generated molecule is synthesizable, and one viable synthesis route can be determined immediately upon molecule generation. Although promising, this model heavily depends on the available chemical synthesis information.

## 5. MOLECULE GENERATION AND OPTIMIZATION BY AUTOENCODER AND VARIATIONAL AUTOENCODER

**Theory of autoencoder (AE) and VAE.** AE maps high dimensional inputs into low dimensional latent variables by encoder and maps the latent variables back to their previous representation by decoder. The goal of AE is to seek the most representative latent variables of the inputs by minimizing the reconstruction loss that characterizes the difference between the input and the output. VAE, proposed in 2013 (Kingma & Welling, 2013), is similar to AE in form but differs from it in part of the underlying theories. In VAE, the encoder maps each input into a latent variable distribution. In many cases, the distribution is characterized by Gaussian parameters such as mean and standard deviation. Different from the traditional variational inference using per-input optimization, the VAE encoding process uses a single parameter set for modeling the connection between all input data and the latent variables (called amortized inference). After encoding, latent variables are sampled from the latent distribution for decoding. VAE optimization objective is generally to minimize both the reconstruction loss and Kullback-Leibler (KL) divergence that characterizes the dissimilarity between distributions. Minimizing KL divergence can be considered as equivalent to maximizing the log-likelihood. The intractability of the associated optimization motivates the use of a tractable log-likelihood lower bound, which leads to the derivation of evidence lower bound. A technique called reparameterization trick is used during the optimization to reduce variance in the gradients by reorganizing the gradient computation. One possible simplified view of VAE is in Figure 3. More details about VAE theory are provided by Doersch (2016) and Kingma & Welling (2019). Many applications of AE and VAE in molecule generation and optimization exist and ten of them are summarized next. Note that many applications involve the modification of encoder and decoder of AE and VAE.

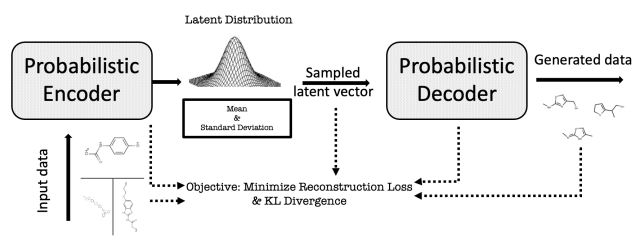

**Figure 3. A simplified example of VAE.**

**Work by Gómez-Bombarelli et al. (2018).** Guided by a sentence generation model combining VAE and RNN (Bowman et al., 2015), Gómez-Bombarelli et al. develop a VAE-based drug design model with an additional predictor. The encoder converts the discrete SMILES representation of molecules to the continuous vector representation while the decoder converts between two representations in the opposite direction. Both RNN and dynamic CNN for sequence modelling are used for the encoder and dynamic CNN is found to perform better than RNN in encoding due to the special substring structure corresponding to some chemical substructures (e.g. functional groups) (Kalchbrenner et al., 2014). RNN with multiple gated recurrent units are used in the decoder. MLP is used in the predictor, which evaluates the properties of molecules corresponding to the latent representations. The continuousness of the latent representation at the VAE bottleneck is necessary because it realizes efficient molecular interpolation, gradient-based optimization, and other exploration tasks such as slightly modifying given chemical structures.

Gradient-based optimization, the most important task, starts with the latent vector, and modifies the latent vector following the direction of potential molecular property improvement before the decoding step. The latent space for optimization is unbounded because molecule space is (almost) boundless. During optimization, Bayesian inference methods can be used to search for the molecules corresponding to the neighborhood of the global optimum. Note that their VAE model could generate invalid SMILES strings. Thus, the authors use RDKit to filter out the invalid SMILES strings after the decoding step.

**GVAE (G: grammar) (Kusner et al., 2017).** GVAE aims to prevent the invalid molecular representations in the previous model by Gómez-Bombarelli et al. Concretely, GVAE doesn't directly operate on SMILES strings. Rather, it represents syntactically valid SMILES strings by parse trees based on a grammar (Socher et al., 2013). The parse trees can capture the structures of SMILES strings and they have high validity. GVAE encodes and decodes directly using the parse trees. As a result, the generated molecular representations by VAE have high validity. In addition, facilitated by the parse tree representation, GVAE only needs to learn semantic properties from the input without the need to learn the syntactic rules, and this could improve its performance.

**VAE with constrained Bayesian optimization (Griffiths & Hernández-Lobato, 2020).** Griffiths and Hernández-Lobato propose another approach to address the issue with generating invalid molecules by the aforementioned model from Gómez-Bombarelli et al. BO used by Gómez-Bombarelli et al. could select molecules corresponding to the undesirable region of the VAE latent space. The authors term such region as the dead region. Dead region corresponds to the molecule categories unseen by VAE during training. The formation of dead region is due to three reasons. First, the latent space of VAE could have high dimensionality. Second, some gap can form in the VAE latent space because some molecular types are unseen during VAE training. Third, some locations in the latent space are highly unlikely under the prior. To address the issue with dead region, the authors add a constraint during BO. Concretely, the probability of successful decoding must exceed a threshold during optimization. After this modification, a higher chemical validity is achieved for the generated molecules.

**GraphVAE (Simonovsky & Komodakis, 2018).** GraphVAE is an early VAE-based molecule generative model that operates directly on graphs. The authors use GNN for graph encoding and they point out that other graph embedding techniques will also work in their application. The main challenge in their task is decoding. The authors propose a novel decoding procedure. During decoding, a point from the VAE latent space is selected and decoded to a probabilistic fully-connected graph represented by a probabilistic adjacency matrix and probabilistic features. Concretely, the adjacency matrix represents the probability of node and edge presence in the graph, and the probabilistic features indicate the probability associated with the identities of nodes and edges. A maximal size of the decoded graph is set to render the computation associated with dense graphs tractable. Note that the decoder is based on MLP, and the decoder itself is deterministic. Another challenge in their task is to determine the reconstruction loss, which requires graph comparison. However, no fixed node ordering is defined for the graphs in their task and the adjacency matrix for graph comparison is affected by node ordering. To address this issue, the authors design a technique for approximate graph matching.

**JTVAE (JT: junction tree) (Jin et al., 2018a).** JTVAE is an early and intricate model operating on molecular graphs and junction trees. To endorse the use of graph-structured representation instead of SMILES strings for molecule representation in their task, the authors point out that SMILES representation is limited for VAE-based generative model because similar molecules with drastically different SMILES strings prevent VAE from learning smooth molecule embeddings. Compared to the aforementioned graph-based molecular generative models, one distinct feature of JTVAE is the use of the coarse molecular representation (i.e. junction tree) in addition to the use of the fine representation (i.e. raw molecular graph with atoms as nodes and chemical bonds as edges). In a JT, each node corresponds to a chemical substructure obtained via a tree decomposition process. The JT structure corresponds to the coarse relative arrangement of the substructures. Generating molecules as JTs has two advantages over the atom-by-atom generation scheme. First, the atom-by-atom generation approach could lead to invalid intermediate molecular graphs while using suitable molecular substructures as nodes in JT can circumvent this issue. Second, using JTs allows generation of larger molecules.

JTVAE has a graph encoder, a tree encoder, a tree decoder, and a graph decoder. A graph message passing network is used for the graph encoder. A tree message passing network with GRU is used for the tree encoder. During the encoding process of the JT associated with an input molecule, messages are iteratively propagated from the leaf nodes to the JT root (called the bottom-up phase). JTVAE decoding process has two stages: JT decoding followed by graph decoding. During the first stage, the JT is decoded from a latent embedding sequentially (i.e. node by node) with the depth-first order. A message is updated by GRU as the JT is traversed. When a node is visited, the updated message, tree embedding, and the node feature are collectively used to predict the presence of a child node. Node label prediction follows immediately if a child node is generated. Feasibility check is also performed during tree decoding. During the graph decoding stage, the decoded JT is encoded again by mostly following the aforementioned tree encoding process, except that the message propagation now has both bottom-up phase and the top-down phase with the opposite propagation direction. The molecular graph is then decoded based on the encoding of the decoded JT, together with other information such as the graph latent embedding. Note that teacher forcing is used during the training of both tree decoder and graph decoder.

To improve the properties of the generated molecules, the authors use Bayesian optimization to guide the generation process. The authors also perform the constrained molecular optimization to find molecules that have improved properties and are similar to the original molecules.

**HierVAE (Hier: hierarchical) (Jin et al., 2020a).** JTVAE with structure-by-structure generation scheme is able to generate larger molecules when compared to most models with atom-by-atom generation scheme. However, there are two hurdles associated with using larger substructures (i.e. motifs) in JTVAE to generate even larger molecules such as polymers. First, combinatorial enumeration is required during the JTVAE decoding process to assemble molecule substructures; thus, using larger motifs in JTVAE is computationally intractable. Second, the substructures in JTVAE are restricted to certain types; however, structures of the larger motifs could be flexible. To overcome these challenges and generate larger molecules, HierVAE, another intricate model with a pair of hierarchal encoder and decoder, is proposed.

HierVAE operates on a system composed with three stacked graphs. The bottom graph is a raw molecular graph capturing the fine molecular details. The middle graph is the attachment graph representing the connectivity between motifs. In the attachment graph, each node corresponds to an attachment configuration, and each edge captures the decoding order of the nodes. The top graph is a motif layer capturing the coarse connectivity of motifs. In the motif graph, each node represents a motif, and each edge represents the overlap between motifs. Three graphs are connected to each other via vertically directed edges for information propagation. Specifically, an atom node in the bottom layer has an outcoming edge to the attachment node if the atom belongs to the chemical structure captured by the attachment node. An attachment node has an outcoming edge to its corresponding motif node.

During encoding, message passing neural network in the bottom graph updates the atom representations. The updated atom information and the attachment node embedding are used to update the attachment node features by MLP. Then, MPNN in the attachment graph updates the attachment node representation. The updated attachment node information and the motif node embedding are used to update the motif node features by MLP. Finally, MPNN in the motif graph updates the motif representations. A distribution of latent embedding is then obtained based on the updated motif information. During each step of decoding, the intermediate hierarchical graph is encoded to obtain intermediate motif and atom representations. The motif and its possible attachments are predicted via MLPs based on the VAE latent embedding and the representation of the motif to which the new motif will be attached. The exact attachment of the new motif is then predicted based on atom pair representations and VAE embedding.

HierVAE performs well in generating large molecules. The authors also broaden the utility of the hierarchical representation to a graph-graph translation model for molecule optimization (Jin et al., 2018b).

**OPTIMOL (Boitreaud et al., 2020).** OTTIMOL is proposed to generate molecules that are specific to a target by incorporating the information about the interaction between the molecules and the target via docking (i.e. a costly computation method to estimate molecule-target interaction). Concretely, a VAE is used to generated molecules, which are then docked. Based on the docking result, the molecules with high affinity toward the target are fed to VAE for fine tuning. OPTIMOL is a closed-loop design that iteratively optimizes the binding affinity of generated molecules. In OPTIMOL, the encoder of the VAE model operates directly on molecular graphs. However, to avoid high computation cost associated with a complicated decoder required for outputting graphs, the authors represent molecules as sequences during decoding. Using SMILES could result in invalid molecules. Thus, the authors use Selfies (Krenn et al., 2019), a recently proposed SMILES alternative, and consequently, 100% valid molecules are generated.

**Molecule CHEF (Bradshaw et al., 2019).** Molecule CHEF is a generative model that also concerns the molecule synthesizability. Note that Molecule CHEF is a Wasserstein Autoencoder (WAE) rather than a VAE (Tolstikhin et al., 2018). However, it is summarized here due to its similarity to VAE in terms of model components (i.e. encoder and decoder). Unlike the aforementioned VAE models encoding a single molecule each time, Molecule CHEF encodes a set of reactant molecules. Concretely, Molecule CHEF encoder uses the gated GNN to determine the representation of each reactant molecule. These representations are summed, and the summation is used to determine the latent distribution of the molecular set via a feed forward network. During decoding, the Molecule CHEF decoder maps the latent representation to a set of reactant molecules via RNN parameterized by the WAE latent representation. Note that all possible reactant molecules to be generated are picked from a fixed molecule reservoir. Next, a reaction prediction model is used to map from the reactant molecule set to the final product molecule. As a result, Molecule CHEF can conveniently move around in the latent space, and pick a reactant molecule set to form the product molecule, of which the synthesis pathway is simultaneously determined.

**Shape-guided molecule generation (Skalic et al., 2019).** Skalic et al. propose perhaps the first molecule generative model guided by molecular shape features. Their model has a shape VAE and captioning networks. The shape VAE is based on CNN and it operates on molecule shape representations. Pharmacophores (abstract and informative molecule representations) are fed to the decoder to preclude the reconstructed shape representations that deviate much from the pharmacophores. The reconstructed molecule shape representations are then fed to captioning networks to generate molecules that fit the shape representations.

**PaccMannRL (Born et al., 2021b).** PaccMannRL is developed to generate molecules that target the transcriptomic profiles of cell lines. As a result, disease context can be incorporated during the molecule generation process. PaccMannRL incorporates a hybrid VAE in an RL framework. The encoder and decoder of the hybrid VAE are from two VAEs. The encoder is from a profile VAE pre-trained with gene expression data, and it can encode gene expression to the latent space. The decoder is from a SMILES VAE pre-trained with SMILES molecular representations, and it can decode the latent representation to generate molecules. The hybrid VAE is retrained to generate molecules given the gene expressions. To generate molecules with efficacy against the biomolecular profiles associated with the gene expressions, the hybrid VAE is incorporated into an RL framework with an anticancer drug sensitivity prediction model (PaccMann) to assign rewards that guide the generation process.

# 6. MOLECULE GENERATION AND OPTIMIZATION BY GENERATIVE ADVERSARIAL NETWORK

**Theory of GAN.** GAN (Goodfellow et al., 2014), proposed in 2014, is a likelihood-free generative model capable of implicitly learning the distribution of real training data. A basic GAN consists with a generator and a discriminator that are often based on deep neural network. Generator generates fake data by mapping its input (random data distribution, e.g. normal distribution) to a higher dimensional space. The discriminator is essentially a binary classifier outputting the probability of its input being real or fake. The training goal of the generator and discriminator are conflicting. The discriminator is trained to maximize its ability of discerning the fake data from the real training data while the generator is trained to maximize its ability of producing fake data undetectable by the discriminator. From the perspective of the game theory, the generator and discriminator are playing a zero-sum (in terms of their losses) noncooperative game and the training process stops when the Nash-equilibrium is reached. The training process can be viewed as a "minmax" optimization problem with the objective to detect fake data from the real ones. The generator tries to minimize the objective term while the discriminator attempts to maximize it. A simplified visual depiction of GAN is in Figure 4. Despite many successful applications of GAN in image, audio, videos and texts related tasks, challenges associated with GAN still exist (e.g. training nonconvergence and low diversity of generated data). The following surveys by Jabbar et al. (2020), Lin et al. (2020), Pan et al. (2019), and Saxena & Cao (2021) provide more details about GAN and its variants. Five applications that involve GAN are summarized next.

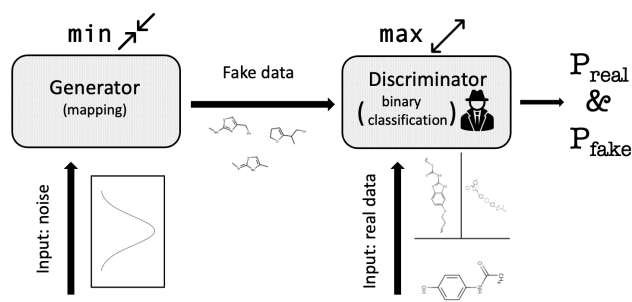

**Figure 4. A simplified view of GAN.**

**MolGAN (Mol: molecular) (De Cao & Kipf, 2018).** MolGAN is developed based on objective-reinforced GAN (ORGAN) designed for strings (Guimaraes et al., 2017). MolGAN is the first graph-based molecule generation model using GAN. Specifically, Wasserstein GAN (WGAN) (Arjovsky et al., 2017), a more stable GAN variant, is used together with a reward network. The generator is based on MLP, and it has an input of standard normal distribution. It generates the node feature matrix and adjacency tensor associated with entire graph all at once for computational efficiency. The discriminator is consisted with a node-order permutation-invariant graph convolutional network variant (relational-GCN by Schlichtkrull et al. (2018)) and MLP for binary classification. Implicitly, GAN promotes the generation of molecules mimicking the training molecules. To promote molecule functionality and novelty, RL is incorporated into the model. Specifically, to help convergency and maximize approximately expected future reward, deep deterministic policy gradient is used. Reward is assigned to generated molecules based on their properties evaluated by an external software. Similar to the discriminator, the reward network also consists with relational-GCN and MLP. To connect the reward network with GAN, the generator is trained with a loss function that is a linear combination of WGAN loss and RL loss. A hyper parameter is used to regulate the trade-off between two losses. To ensure high validity of the generated molecules, zero reward is assigned when invalid molecules are generated. Note that the generated molecules from MolGAN have low diversity due to model collapse, one common GAN issue, despite careful tunning. Also, the one-shot generation scheme (i.e. output the molecular graph at once) in MolGAN deteriorates model performance when handling large molecular graphs. Thus,

MolGAN is only appliable toward graphs with a small number of nodes. Consequently, only small molecules containing up to nine heavy atoms (i.e. carbon, nitrogen, oxygen, and fluorine) are used to train the discriminator of MolGAN.

**L-MolGAN (L: large) (Tsujimoto et al., 2021).** L-MolGAN is proposed to address the incapability of MolGAN to effectively generate large molecular graphs. The authors of L-MolGAN point out that large graphs generated by MolGAN are likely to be disconnected. Thus, they make a simple change to MolGAN. Concretely, during training, depth-first search is used by L-MolGAN to evaluate the connectivity of generated molecular graphs. A zero reward is assigned when a disconnected molecular graph is generated. As a result, large and connected molecular graphs can be effectively generated by L-MolGAN.

**GAN with adaptive training (Blanchard et al., 2021).** The work by Blanchard et al. is another follow-up work of MolGAN. Specifically, it addresses the model collapse issue of MolGAN by adaptive training that involves genetic algorithm. Genetic algorithm is inspired by the natural selection process. Genetic algorithm is often used to solve optimization and search problems (Mirjalili, 2019; Parrill, 1996; Whitley, 1994). Example operators of genetic algorithm are mutation, crossover, and selection. This work uses crossover that generates a child entity by combining the information of two parent entities. Adaptive training starts with using GAN to generate novel molecules and store them during a training interval. Then, some of the training molecules are replaced by the generated novel molecules and the adaptive training process repeats. The crossover operator of genetic algorithm is used to recombine the training molecules and generated novel molecules, and some of the resulted molecules are also used to replace the training molecules.

**Mol-CycleGAN (Maziarka et al., 2020).** Mol-CycleGAN is a GAN-based model that optimizes molecules while keeping them similar to the starting molecules. As mentioned earlier, one benefit of keeping molecular similarity is to mimic the practical lead optimization process in drug development. In addition, the authors point out that the synthesizability of optimized molecules can be improved if they are similar to the starting practical molecules. Mol-CycleGAN is specifically based on the recent method Cycle-GAN (Zhu et al., 2020), and parameters balancing between the property improvement and similarity is used in the model for trade-off.

**Work by Méndez-Lucio et al (2020).** Méndez-Lucio et al. develop a GAN-based model conditioned on transcriptomic data for generating active molecules that can induce desired transcriptomic profiles. Their model has two stages involving two conditional GANs, of which the second one refines the results from the first one. In addition, an autoencoder is pre-trained so that its decoder can map from latent representations to molecules. During the first stage, the gene expression signature and random noise are fed to the first GAN and its generator produces a molecular representation, which is compatible with the decoder of AE. Then, the discriminator of the first GAN determines whether the molecular representation corresponds to a real molecule. Meanwhile, a conditional network decides if the molecular representation matches the gene expression signature. During the second stage, the molecular representation and the gene expression signature are fed to the second GAN and the same process repeats.

# 7. MOLECULE GENERATION AND OPTIMIZATION BY FLOW

**Theory of flow.** Flow was proposed in the year 2010, and it has been widely used since 2015 (Dinh et al., 2014; Rezende & Mohamed, 2015; Tabak & Vanden-Eijnden, 2010). Different from VAE and GAN, flow is capable of explicitly learning the distribution from the input data. In its latent space, flow often aims to transform a simple distribution (e.g. a normal distribution) to a highly complex one (e.g. a multi-modal distribution) through an invertible and differentiable transformation. A single function is usually insufficient for a complex transformation. Thus, a series of mappings (i.e. function composition) are typically used with the change of variable theorem since function composition maintains invertibility. Data is generated through the inversed transformation of latent variable. Loss directly determined using model input and output is used for optimization. One possible simplified view of flow is in Figure 5. Practical design considerations of flow are high expressivity (i.e. capture data distribution) and efficient computation (e.g. fast computation of the Jacobian matrix determinant). Examples of flow models include linear flow, autoregressive flow, residual flow, and infinitesimal flow. Detailed flow theory can be found in papers by Kobyzev et al. (2020), Papamakarios et al. (2019), and Weng (2018). Five applications of flow are summarized next.

**GraphNVP (NVP: non-volume preserving) (Madhawa et al., 2019).** GraphNVP is the first model that leverages the invertible flow in molecular graph generation. The authors claim that the exact likelihood maximization of flow is essential in designing molecules, of which the properties could be altered drastically by minor perturbation such as changing one atom. Invertible flow is commonly used to generate images, of which the grid structure differs from the molecular graph structure, which is highly sparse due to chemical valence. To render flow appliable to sparse graph structures, the authors transform the adjacency tensor and feature matrix associated with the molecular graph into the latent representation via two types of reversible affine coupling layers (i.e. adjacency coupling layers and node feature coupling layers) based on the real-valued non-volume preserving transformations (Dinh et al., 2016). During the forward transformation, the change of variable formula can't be directly used for discrete data distribution. Therefore, they use dequantization to convert the discrete data distribution to a continuous distribution. Specifically, uniform noise is added to the discrete adjacency tensors and feature matrix for dequantization. To efficiently generate valid molecular graphs, GraphNVP follows a two-step generation scheme. Adjacent tensor is generated first, and the feature matrix is generated next based on the adjacent tensor.

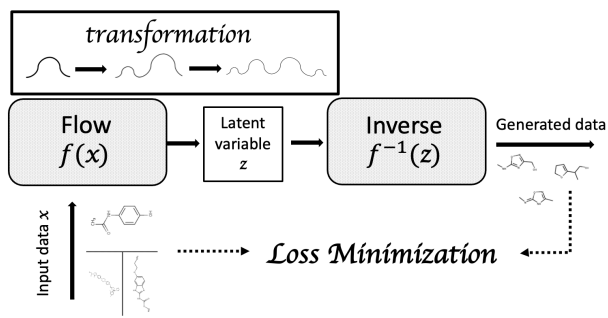

**Figure 5. A simplified example of flow.**

**GRF (graph residual flow) (Honda et al., 2019).** In one follow-up work of GraphNVP, Honda et al. argue that the coupling flow is still unsuitable for sparse graph structures because only one node representation can be updated in each mapping layer of the coupling flow. Thus, a large number of layers are required to update representations of many nodes once. They suspect that the performance of GraphNVP is compromised by the subpar mapping effectiveness, which cannot be remedied by further development of the partition-based coupling flow. This hypothesis motivates them to leverage residual flow, which is more flexible and complex than the coupling flow and doesn't depend on the variable partition (Chen et al., 2019). Their model, consisting of residual flow and generic graph convolutional network, is called graph residual flow (GRF). GRF is capable of modifying all node attributes in each mapping layer. The authors demonstrate that GRF requires fewer trainable parameters than does GraphNVP to reach similar generation performance. In addition, they mathematically derive conditions to ensure that GRF is invertible during training and sampling. Same as GraphNVP, GRF follows a one-shot generation scheme.

**GraphAF (AF: autoregressive flow) (Shi et al., 2020b).** Despite high efficiency, typical one-shot generation scheme cannot ensure molecule validity, and fully capture graph structures. GraphAF, based on autoregressive flow (Papamakarios et al., 2017), sequentially generates nodes and edges. It extracts the intermediate sub-graph information by GNN, and it ensures validity through valency checking. RL is incorporated into GraphAF to guide the generation for molecular property optimization. The efficiency of GraphAF training process is improved via parallel computing.

**MoFlow (Zang & Wang, 2020).** MoFlow is a one-shot generation model that maintains the efficiency of one-shot generation scheme while ensuring chemical validity. MoFlow follows three steps. Different chemical bonds (e.g. single and double bonds) represented by multi-type edges are generated first by a variant of the Glow model (Kingma & Dhariwal, 2018). Next, the atoms are generated by a graph conditional flow based on the bond information with the help of graph convolutions. The bonds and atoms are subsequently arranged into a molecular graph. Validity of the graph is ensured by considering the bond-valence constraints.

**GraphDF (DF: discrete flow) (Luo et al., 2021).** Issues with the dequantization process in the previous models such as GraphNVP motivate the development of GraphDF. Recall that dequantization is used to convert discrete data distribution to continuous distribution because the change of variable formula doesn't apply to discrete data distribution during the forward transformation. The authors of GraphDF claim that the dequantization process results in an imprecise representation of the discrete distribution associated with the graph structures, and the dequantization process leads to difficult training and compromises model performance. GraphDF circumvents the dequantization process. GraphDF is a discrete latent variable model using invertible modulo shift transformations to map discrete latent variables to nodes and edges during the molecular graph generation process.

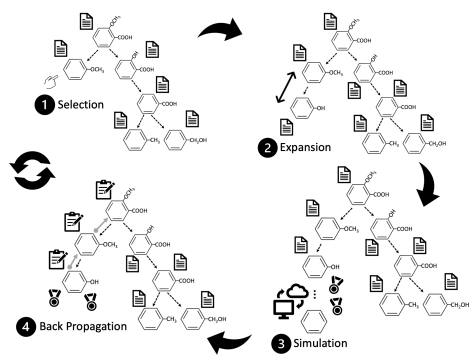

**Figure 6. A visual depiction of MCTS.** Details of four steps (i.e. selection, expansion, simulation, and backpropagation) are summarized in Section 8.

# 8. MOLECULE GENERATION AND OPTIMIZATION BY MONTE CARLO TREE SEARCH

**Theory of MCTS.** MCTS was proposed in 2006 (Coulom, 2006). It is a combination of Monte Carlo method, RL, and tree search. Monte Carlo method solves deterministic problems through repeated random sampling processes (Rubinstein & Kroese, 2016). MCTS improves upon the brutal-force tree search by leveraging Monte Carlo method to determine the most promising moves. Implemented with neural networks, MCTS played a central role in improving the computer Go (Silver et al., 2016). MCTS repeats four steps: selection, expansion, simulation, and backpropagation. The selection process starts from the root of search tree, and picks successive child nodes by following a tree policy until a leaf node is reached. Tree policy is used for the selection of child nodes leading to promising moves. Tree policy should balance exploration and exploitation. Thus, upper confidence bound algorithm is commonly applied in tree policy. Expansion occurs after selection via adding one or more child nodes to the selected leaf node. Simulation follows expansion via the rollout algorithm, which iteratively simulates multiple trajectories from the expanded state before a terminal state is reached. Rollout algorithm is determined by the roll out policy, which usually uses random process due to its simplicity for fast computation. During the simulation, reward associated with the expanded node is accumulated and used for the last step of each MCTS cycle: the backpropagation. In backpropagation, the reward is back propagated from the expanded node to the root node for information update. MCTS cycle repeats until no time remains or computation is exhausted. A visual depiction of MCTS is in Figure 6. Despite being more efficient than the brutal-force search, MCTS is still slow and requires a large amount of memory. More MCTS details are in papers by Browne et al. (2012) and Swiechowski et al. (2021). Two applications of MCTS are summarized next.

**UnitMCTS (Rajasekar et al., 2020).** UnitMCTS is a follow up work of the aforementioned MolDQN model in Section 4. UnitMCTS makes one unit change to the molecule during each step. The possible changes include atom addition, bond addition, bond removal, and bond replacement. After the MCTS simulation step, the property of the final molecule is evaluated with the help of a property evaluator and used for the backpropagation step. Like MolDQN, UnitMCTS doesn't require training dataset. In addition, the authors perform constrained molecule optimization that prevents modified molecules from deviating much from the staring molecules. Concretely, during the MCTS expansion step, molecules that are added to the search tree must be similar to the starting molecules.

**RationaleRL (Jin et al., 2020b).** RationaleRL is designed for generating molecules that satisfy multiple property requirements by using rationales, which are small molecular substructures possibly responsible for the desired chemical properties in the molecules. RationaleRL first identifies single-property rationales from molecules with desired properties. Then, single-property rationales can be combined to form multi-property rationales (i.e. substructures possibly responsible for several desired chemical properties). Last, the rationales are expanded to molecular graphs through graph completion processes.

MCTS is used in RationaleRL to extract the single-property rationales. During the rationale extraction process, the root of the search tree is the molecular graph with the desired property. Each state in the search tree is a subgraph of the root graph. Subgraphs are obtained by deleting the peripheral bond(s) of the root graph. A property predictor aids the MCTS process to identify the subgraph that represents the rationale with the desired chemical property.

# 9. DISCUSSION, CHALLENGES, AND CONCLUSION

**Molecular property prediction.** Approaches for computational molecular property prediction are mainly based on domain knowledge, data, and the mix of the two. Recently, data-driven approach centers on DL and specifically GNN. This survey includes DL theory and an in-depth summary of many GNN applications in molecular property prediction. A comprehensive list of GNN applications in this subfield can be found in the review by Wieder et al. (2020).

There are some specific tasks in this subfield, and a few of them are summarized here. A common task is to capture molecular global information for property prediction by GNN. Methods designed for this task include adding global information to atom features, using the precomputed global molecular feature by other software, and using an extra global node (Coley et al., 2017; Li et al., 2017; Yang et al., 2019a). Predicting the interaction between two molecular entities (including protein) is another common task. Methods include message passing between two graphs and link prediction of a molecular network (Jiang et al., 2020; Wang et al., 2021b; Zitnik et al., 2018). Dai et al. (2021), Feinberg et al. (2018), and Lin (2020) also propose methods for this task. A third task is to check if graph representation indeed leads to superior performance of molecular property prediction compared to other representations, and some related work has been done by Jiang et al. (2021).

There are several major hurdles in molecular property prediction using GNN. The most challenging one is perhaps the low availability of molecular data for some specific prediction tasks, despite the increased amount of general property data and improved high-throughput technologies. DL is extremely data consumptive. Methods such as transfer learning and meta-learning are proposed to partially alleviate this issue (Altae-Tran et al., 2017; Hu et al., 2019; Nguyen et al., 2020). Yet, it is challenging to rely mostly on theoretical development of models to overcome the problem of low data availability. Combining active learning and efficient data acquisition instruments might be a promising direction (Settles, 2009). Besides low data availability, a fraction of data with low accuracy potentially due to instrument errors or human errors can also drastically compromise the predictor performance. Leveraging AI-based models to detect these inaccurate data will be helpful, but it is also challenging. In addition, uncertainty quantification of the predictions is another challenge (Smith, 2013), and some related work has been done by Hirschfeld et al. (2020). One more hurdle for GNN-based molecular property prediction is to consider molecule chirality. One approach is proposed to take the tetrahedral chirality into account during chemical property prediction by modifying the aggregation function of GNN (Pattanaik et al., 2020). More methods for other chirality types can be developed (Clayden et al., 2012).

**Molecule generation and optimization.** AI-based molecule generation and optimization provide a possible way to navigate in the vast unexplored chemical space. Section 3 to Section 8 provide a detailed summary of 29 examples involving deep generative models (e.g. RNN, VAE, GAN, and flow) and RL (including MCTS) for molecule generation and optimization. Moreover, attributes and highlights of 60 additional applications involving these models for the same purpose are briefly summarized in Table 3. Other techniques such as Bayesian optimization, genetic algorithms, evolutionary algorithms, and energy-based methods are also used recently as a major part of some models developed for molecule generation and optimization (Ahn et al., 2020; Bagal et al., 2021; Hataya et al., 2021; Korovina et al., 2020; Kwon & Lee, 2021; Leguy et al., 2020; Liu et al., 2021a; Nigam et al., 2019; Nigam et al., 2021; Shen et al., 2020). Given the abundance of the models and their variants, frameworks such as GuacaMol and molecular sets (MOSES) are developed for comparing the model performance (Brown et al., 2019; Polykovskiy et al., 2020), and Rigoni et al. (2020a) systematically compare many models for molecule generation and optimization. In addition, a python package is developed by Reeves et al. (2020) specifically for molecule generation.

One common task in molecule generation and optimization is to maintain the scaffold during the process. Solutions to handle this task include constrained optimization, multi-objective optimization, cycle GAN, etc. (Maziarka et al., 2020; Zhou et al., 2019). Another task is to effectively generate larger molecular graphs, and models such as HierVAE and L-MolGAN are designed for this purpose (Jin et al., 2020a; Tsujimoto et al., 2021). A third common task is to generate drugs with the desired interaction property toward the given target, and molecular simulation can be used to help with this task (Boitreaud et al., 2020).

The biggest obstacle in AI-based molecule generation and optimization perhaps also centers on the low availability of some molecular data. Methods used to partially alleviate this challenge in molecule generation and optimization include transfer learning, semi-supervised learning, and self-training (Amabilino et al., 2020; Kang & Cho, 2018; Segler et al., 2018a; Yang et al., 2020a). Ensuring high synthesizability of generated or optimized molecules is another challenge. An RL-based model using chemical reactions as state transitions and a WAE operating on a group of reactant molecules are proposed to overcome this challenge (Bradshaw et al., 2019; Horwood & Noutahi, 2020). Note that there is a field that is pertinent to the topic of this survey and focuses specifically on AI-based chemical synthesis (e.g. chemical reaction prediction and retrosynthetic analysis) (Do et al., 2019; Mo et al., 2021; Segler et al., 2018b; Shi et al., 2020a; Wang et al., 2020). Another challenge lies within 3-D molecule generation and optimization. Graph-based and string-based methods are not suitable to capture 3-D molecular details, which

**Table 3. Attributes and highlights of additional applications in molecule generation and optimization.**

| Applications | Attributes and highlights |
|---|---|
| Work by Olivecrona et al. (2017) | An early SMILES-based molecule generation model using RNN in an RL framework. |
| ORGAN by Guimaraes et al. (2017) | An early model based on <u>**GAN**</u> for <u>**O**</u>bjective-oriented SMILES molecule generation enabled by <u>**RL**</u>. |
| ORGANIC by Sanchez-Lengeling et al. (2017) | A follow-up work of <u>**ORGAN**</u> for <u>**I**</u>nverse-design <u>**C**</u>hemistry. |
| DruGAN by Kadurin et al. (2017) | A model based on an adversarial AE. |
| DiPol-GAN by Guarino et al. (2017) | A <u>**GAN**</u>-based model that uses <u>**Di**</u>fferentiable <u>**Pool**</u>ing, and is able to capture the hierarchical molecular representations. |
| ReLeaSE by Popova et al. (2018) | An RNN-based model that is in a <u>**Re**</u>inforcement <u>**Lea**</u>rning framework for <u>**S**</u>tructural <u>**E**</u>volution, and has improved performance compared to the related models. |
| RANC by Putin et al. (2018a) | Full name: <u>**R**</u>einforced <u>**A**</u>dversarial <u>**N**</u>eural <u>**C**</u>omputer. A model that is related to GAN and RL, and operates on SMILES representations. |
| ATNC by Putin et al. (2018b) | Full name: <u>**A**</u>dversarial <u>**T**</u>hreshold <u>**N**</u>eural <u>**C**</u>omputer. A work related to RANC. |
| Work by Lim et al. (2018) | A model that is based on conditional VAE, uses SMILES representations, and is able to control several properties of the designed molecules simultaneously. |
| Work by Ma et al. (2018) | A model based on VAE with a regularization framework for ensuring molecule sematic validity. |
| VJTNN by Jin et al. (2018b) | Full name: Variational Junction Tree Encoder-Decoder. A model that is trained by molecular pairs and can map from one graph to another with improved properties. |
| CGVAE by Liu et al. (2018) | A graph-based model using a <u>**C**</u>onstrained <u>**G**</u>raph <u>**VAE**</u>, of which the decoder assumes sequential graph extension steps. |
| Work by Kang & Cho (2018) | A model that leverages unlabeled molecules by using a semi-supervised VAE, of which the training molecules are only partially labeled. |
| Work by Li et al. (2018) | A graph-based model that uses RNN and GNN, and expresses probabilistic relationships among nodes and edges of graphs. |
| DcGAN by Bian et al. (2019) | A model that uses <u>**D**</u>eep <u>**c**</u>onvolutional <u>**GAN**</u> for designing molecules targeting cannabinoid receptors. |
| ALMGIG by Pölsterl & Wachinger (2019) | Full name: <u>**A**</u>dversarial <u>**L**</u>earned <u>**M**</u>olecular <u>**G**</u>raph <u>**I**</u>nference and <u>**G**</u>eneration. A GAN-based model that circumvents the need for determining reconstruction loss explicitly. |
| LatentGAN by Prykhodko et al. (2019) | A model combining AE and GAN. |
| Work by Sattarov et al. (2019) | A model using generative topographic mapping to explore the VAE latent space. |
| Work by Mansimov et al. (2019) | A molecule conformation (i.e. geometry) generation model based on VAE and GNN. |
| Work by Bresson & Laurent (2019) | A model using VAE with a two-step decoder: 1. atom generation and 2. assembling. |
| Tiered VAE by Chang (2019) | A model that is based on <u>**VAE**</u> and GNN, and uses <u>**Tiered**</u> latent representations (i.e. atom level, group level, and molecule level) of molecular graphs to capture functional groups. |
| Work by Hong et al. (2019) | A model with an adversarially regularized AE. |
| MHG-VAE by Kajino (2019) | A <u>**VAE**</u>-based model using <u>**M**</u>olecular <u>**H**</u>ypergraph <u>**G**</u>rammar to encode chemical constraints for ensuring validity of generated molecules. |
| RL-VAE by Kearnes et al. (2019) | A graph-to-graph <u>**VAE**</u>-based model that leverages <u>**RL**</u> during decoding. |
| Work by Kwon et al. (2019) | A model that uses RL and an improved version of VAE for one-shot graph generation. |
| DR-AIM by Zhang et al. (2019a) | A model using a <u>**D**</u>eep <u>**R**</u>einforced framework with <u>**A**</u>dversarial <u>**I**</u>mitation and <u>**M**</u>ultitask learning (i.e. jointly multi-objective optimization). |
| Work by Jensen (2019) | A work that uses and compares MCTS and genetic algorithm. |
| Work by Pang et al. (2020) | A VAE-based model that learns a latent space energy-based prior model to improve validity of generated molecules represented by SMILES. |
| LFM by Podda et al. (2020) | Full name: <u>**L**</u>ow-<u>**F**</u>requency <u>**M**</u>asking. A SMILES-based model that uses RNN and VAE for fragment-based molecule generation. |
| Modof by Chen et al. (2020b) | Full name: <u>**Mod**</u>ifier with <u>**o**</u>ne <u>**f**</u>ragment. A graph-based VAE model that identifies and modifies fragments of molecules to improve their properties. |

**Table 3 continued…**

| | |
|---|---|
| CogMol by Chenthamarakshan et al. (2020) | A VAE-based model for **Co**ntrolled **g**eneration of **Mol**ecules with high affinity to viral proteins and off-target selectivity. |
| Work by Court et al. (2020) | A VAE-based model for generating 3-D inorganic crystal structures. |
| DeLinker by Imrie et al. (2020) | A VAE-based model linking two fragments to design a molecular graph while leveraging the 3-D structural information. |
| CORE by Fu et al. (2020) | A VAE-based model generating molecules by using part of the input molecular graph (**CO**py) or use substructures from a large substructure reservoir (**RE**fine). |
| Work by Lim et al. (2020) | A VAE-based model designing molecules by taking a desired molecular scaffold graph and adding atoms and bonds to it sequentially. |
| QMO by Hoffman et al. (2020) | A VAE-based model with a generic **Q**uery-based **M**olecule **O**ptimization framework. |
| CCGVAE by Rigoni et al. (2020b) | **C**onditional **C**onstrained **G**raph **VAE** using histograms of atom valences for molecule generation. |
| ChemoVerse by Singh et al. (2020) | A VAE-based model of which the latent space is optimized via a manifold traversal with heuristic search. |
| Work by Tripp et al. (2020) | A VAE-based model of which the latent space is optimized in a sample-efficient manner by weighed retraining. |
| Work by Domenico et al. (2020) | An RNN-based model optimizing multiple properties of molecules in a pair-based manner. |
| DeepGraphMolGen by Khemchandani et al. (2020) | An RL-based model generating molecules binding only to the target of interest (high selectivity) by using multi-objective optimization. |
| SAMOA by Langevin et al. (2020) | An RL-based model for scaffold constrained molecular generation. |
| Work by Simm et al. (2020a) | An RL-based model for 3-D optimization of molecules via a new actor-critic architecture. |
| MNCE-RL by Xu et al. (2020) | An **RL**-based model for molecular optimization with **M**olecular **N**eighborhood-**C**ontrolled **E**mbedding grammars. |
| Work by Polykovskiy et al. (2020) | A model based on an entangled conditional adversarial AE. |
| Work by Ragoza et al. (2020) | A model that uses GAN and VAE, and generates 3-D molecular structures via atomic density grids. |
| Work by Yang et al. (2020a) | A work that uses self-training approach (i.e. select generated molecules to augment targets) to address the low data availability issue. |
| Work by Yang et al. (2020b) | A molecular design model with a massively parallel MCTS algorithm to speed up the computation. |
| MPGVAE by Flam-Shepherd et al. (2021) | A model that is modified based on Graph**VAE**, and includes **MP**NN. |
| DEVELOP by Imrie et al. (2021) | Full name: **DE**ep **V**ision-**E**nhanced **L**ead **OP**timization. A model that is modified based on DeLinker by including CNN, and is designed to incorporate 3-D pharmacophoric constraints for molecule generation. |
| Work by Born et al. (2021a) | A model that involves VAE and RL, and generates molecules targeting a given protein. |
| MoLeR by Maziarz et al. (2021) | Full name: **Mo**lecule-**Le**vel **R**epresentation. A model that uses VAE, and designs molecules by adding fragments or atoms to scaffolds. |
| CMG by Shin et al. (2021) | Full name: **C**ontrolled **M**olecule **G**enerator. A model that uses AE and RNN, and uses sequence translation to handle multi-property molecular optimization. |
| LA-CycleGAN by Wang et al. (2021a) | A model built by embedding **L**STM and **A**ttention mechanism in **CycleGAN**. |
| MolAICal by Bai et al. (2021) | A model that involves GAN and genetic algorithm, and designs drugs considering the 3-D protein pocket being targeted. |
| Work by Li et al. (2021) | A model with efficient quantum GAN. |
| MGRNN by Lai et al. (2021) | Full name: **M**olecular **G**raph **RNN**. A model for molecular structure generation. |
| Work by Mahmood et al. (2021) | A graph-based model that involves RNN and GNN, and captures distribution of nodes and edges conditioned on the remaining graph via a masking technique. |
| SBMolGen by Ma et al. (2021) | Full name: **S**tructure-**B**ased **Mol**ecular **Gen**erator. A SMILES-based model that uses RNN and MCTS, and considers the 3-D structure of the target protein during generation. |
| MolGrow by Kuznetsov & Polykovskiy (2021) | Full name: **Mol**ecular **Gra**ph **Fl**ow. A flow-based model for hierarchical graph generation via recursively splitting each node of the graph into two nodes. |

are essential in some tasks. To overcome this challenge, suitable molecular representations to capture 3-D molecular information are developed (Polykovskiy et al., 2020; Simm et al., 2020b). The last challenge is to incorporate the disease content (i.e. gene information) during molecule generation or optimization. This motivates the development of one model based on a hybrid VAE and another model based on two conditional GANs and an autoencoder (Born et al., 2021b; Méndez-Lucio et al., 2020).

**Conclusion**. By providing a detailed summary of many applications of AI in drug design, this survey helps the readers understand the AI-based models (i.e. GNN, RNN, VAE, GAN, flow, RL, and MCTS) and their variants for handling specific tasks in drug design. Through the discussion about the specific tasks, challenges, and the potential solutions in AI-based drug design, the research directions in this field become evident. The major AI technique in drug design is DL. DL itself is relatively new and so does the field of AI-based drug design. Despite several challenges exist in this field, AI has a large potential to effectively explore the vast chemical space. Further research efforts are required to make AI a powerful tool in drug discovery and development.

## ACKNOWLEDGMENTS

The author would like to thank his research supervisors: Dr. Jeff Calder, Dr. Daniel Boley, and Dr. Maria Gini for advancing his understanding of graph neural network, graph theory, and artificial intelligence.